\documentstyle[aps,epsf,rotate,multicol]{revtex}

\begin{document}

\draft

\title{Self-organized criticality and interface depinning transitions}

\author{Kent B{\ae}kgaard Lauritsen$^{1,2}$\cite{KBL} and Mikko J. 
Alava$^{3,4}$\cite{MJA}}

\address{$^1$Niels Bohr Institute, Center for Chaos and Turbulence Studies,
             Blegdamsvej 17, 2100 Copenhagen, Denmark \\
         $^2$Danish Meteorological Institute, Atmosphere Ionosphere
             Research Division, 2100 Copenhagen, Denmark \\
         $^3$NORDITA, Nordic Institute for Theoretical Physics,
	     Blegdamsvej 17, 2100 Copenhagen, Denmark \\
         $^4$Helsinki University of Technology, Laboratory of Physics,
             02105 HUT, Finland}

\date{\today; submitted: October 15, 1998}
%\date{\today}

\maketitle

\begin{abstract}
We discuss the relation between self-organized criticality
and depinning transitions by mapping sandpile models to equations
that describe driven interfaces in random media. 
This equivalence yields a continuum description
and gives insight about various
ways of reaching the depinning critical point:
slow drive (self-organized criticality),
fixed density simulations, tuning the interface velocity (extremal drive
criticality), or tuning the driving force. 
We obtain a scaling relation
for the correlation length exponent for sandpiles.
\end{abstract}

\pacs{PACS numbers: 64.60.Lx, 05.40.+j, 64.60.Ht, 05.70.Ln}

\begin{multicols}{2}

\section{Introduction}

Many systems respond to external perturbations by avalanches
which behave intermittently with a power-law distribution of sizes.
The paradigm of such self-organized
critical (SOC) behavior is the so-called sandpile model \cite{btw}.
It maintains by an infinitely slow drive a
critical steady-state, where the internal dissipation
balances the external drive. 
Candidates for such phenomena
include granular piles \cite{held-etal:1990,frette-etal:1996},
microfracturing processes \cite{petri-etal:1994},
and earthquakes \cite{gutenberg-richter:1956}.
Despite many theoretical and numerical investigations a thorough
understanding of self-organized criticality is still lacking
\cite{tang:1988,zhang:1989,kadanoff-etal:1989,hwa-kardar:1992,%
paczuski-boettcher:1996,dickman-etal:1998,dhar:1998,%
vespignani-zapperi:1997}.
Fundamental problems which need to be solved involve deriving a
continuum theory which would for instance determine
the upper critical dimension, above which mean-field theory applies
\cite{tang:1988,ZLS:1995}.

Similar behavior can be found from elastic interfaces driven through
random media 
\cite{narayan-fisher:1993,nattermann-etal:1992,leschhorn:1993,laszlo}.
They undergo a continuous (critical) 
depinning transition as the external driving force is varied.
With increasing force one passes from
a phase where the interface is pinned to a depinned phase where the
interface moves with a constant velocity.
Close to the critical point, the motion of the interface takes
place in ``bursts'' with no characteristic size 
and the interface develops scaling described by critical exponents.
These phenomena can be met in 
fluids driven through porous media \cite{rubio-etal:1989}, in
domain walls in magnets (the Barkhausen effect) \cite{zapperi-etal:1998},
in flux lines in type II superconductors \cite{field-etal:1995},
and in charge-density waves \cite{middleton-fisher:1993}.

In this paper we investigate the connections between
self-organized criticality and depinning transitions
\cite{tang:1988,paczuski-boettcher:1996,narayan-fisher:1993,%%%%
zapperi-etal:1998,narayan:1994,paczuski-etal:1996,cule:1998,allau:1999}.
We first establish a generic, exact relation \cite{allau:1999}
between sandpile models and driven interfaces which builds upon previous investigations
of e.g.\ a charge-density wave model \cite{narayan:1994}
and a rice-pile model \cite{paczuski-boettcher:1996}.
Specifically, we discuss the Bak-Tang-Wiesenfeld (BTW) \cite{btw} model
and, as an example, a stochastic sandpile model
\cite{christensen-etal:1996,AL:1996}
through a mapping to a model for interface depinning
with slightly different noise terms.

The mapping enables one to understand the 
slow-drive criticality used in sandpile simulations in terms of standard
concepts for driven interfaces. 
Using the continuum theory for interface depinning
it follows for these sandpile models that the upper critical dimension
$d_c$ is 4, and the relevant noise is of quenched type. The connection
with interfaces allows us to establish a scaling relation for the correlation 
length exponent for sandpile models. In addition, we discuss in the 
interface representation sandpiles driven at fixed density, 
driven at boundaries, and extremal drive criticality.

\section{Sandpiles}

The sandpile models are here defined as follows: to each site of a
$d$-dimensional
lattice (square in $d=2$) of size $L^d$ is associated a variable $z_x$ which
counts the number of grains on that site.
When the number of grains on a site exceeds a critical threshold
$z_x > z_c$, the site is active and it topples.
This means that $2d$ grains are removed from that site and
given to the $2d$ nearest neighbors (nn):
		$z_x \to z_x - 2d$, 
		$z_{nn} \to z_{nn} + 1, ~ \forall nn$.
Sandpiles are usually open such that grains which topple out of the system 
are lost (in one dimension: $z_0 \equiv z_{L+1} \equiv 0$). It is also possible,
as discussed later, to use periodic boundary conditions.
When there are no more active sites in the system,
one grain is added to a randomly
chosen site, $z_x \to z_x + 1$.
The time and number of topplings till the system again contains no active
sites define an avalanche and its internal lifetime and size.
For the BTW model one has $z_c=2d-1$, whereas for stochastic sandpile
models \cite{christensen-etal:1996,AL:1996}
the threshold $z_c$ is not constant.
Below we will focus on 1) the BTW model and 2) a stochastic model
where the threshold $z_c$ is randomly chosen to be for example
$2d-1$ or $2d$ after each toppling, i.e.\ $P(z_c)$, the probability
distribution of the $z_c$'s, is any reasonable choice (i.e.\
decaying sufficiently fast).

In terms of the internal avalanche time, the external drive
is infinitely slow \cite{finite-drive}.
After a transient, the system reaches a steady-state in 
which the slow drive and the dissipation of grains balance each other.
The boundary conditions (BCs) are essential to obtain 
criticality and they are usually of the Dirichlet type, $z\equiv 0$,
such that particles are dissipated to the outside \cite{ZLS:1995}. 
Alternatively, the SOC steady state can be reached by using
bulk dissipation and, e.g., periodic BCs \cite{vespignani-zapperi:1997}.

In the SOC steady-state the probability to have avalanches of lifetime $t$
and size $s$ follow power-law distributions:
$
	p(t) = t^{-\tau_t} f_t(t/L^z)
$
and
$
	p(s) = s^{-\tau} f(s/L^D)  ,
$
with $s\sim t^{D/z}$ and $z(\tau_t-1)=D(\tau-1)$ \cite{Teb99}.
Here the size scales as $s\sim \ell^D$
and the (spatial) area as $\ell^{d}$ (for compact avalanches)
with $\ell$ the linear dimension.
The fact that each added grain will perform of the order of $L^2$
topplings before leaving the system leads to the fundamental result
\begin{equation}
        \left< s \right> \sim L^2
			\label{eq:<s>}
\end{equation}
independent of dimension \cite{tang:1988,kadanoff-etal:1989}.
Thus, $\tau=2-2/D$ and $\tau_t = 1 + (D-2)/z$.
Equation~(\ref{eq:<s>})
yields $\gamma/\nu=2$, where $\gamma$ describes the divergence 
of the susceptibility 
(bulk response to a bulk field) near a critical point,
$\chi = \left< s \right> \sim |\Delta|^{-\gamma}$,
and $\nu$ is the (spatial) correlation length exponent,
$\xi \sim|\Delta|^{-\nu}$ \cite{tang:1988}.
Here $\Delta=\zeta-\zeta_c$ is the control parameter,
$\zeta = \left< z_x \right>$, and the critical value
$\zeta_c = \left< z_x \right>_{\rm SOC}$, where this average
is taken in the slowly driven SOC steady-state with $\Delta=0$
\cite{tang:1988,dickman-etal:1998}.

\section{Interface depinning}

For driven interfaces in random media critical scaling is obtained with a 
force $F$ close to a critical value $F_c$. Depinned
interfaces move with a velocity $v\sim f^\theta$,
with $f=F-F_c \geq 0$. Pinned interfaces
are blocked by pinning paths/manifolds which arise from the
quenched disorder environment. Close to criticality, correlations scale
as $x^{2\chi}$, with $\chi$ the roughness exponent, up to a
correlation length $\xi \sim |f|^{-\nu}$.
The characteristic time scale is $\xi^z$ with $z$ the dynamic exponent
and it follows that $\theta=\nu(z-\chi)$
\cite{narayan-fisher:1993,nattermann-etal:1992}.
Near the depinning transition, the simplest choice
to describe the dynamics of the interface 
is the following continuum equation
('quenched Edwards-Wilkinson', or linear
interface model, LIM)
\cite{narayan-fisher:1993,nattermann-etal:1992,leschhorn:1993}:
\begin{equation}
  \frac{\partial H}{\partial t} =  \nabla^2 H + \eta(x,H) + F.
                                        \label{eq:qEW}
\end{equation}
Here, $H(x,t)$ measures the height of a given site $x$ at time $t$.
The quenched noise $\eta(x,H)$ has correlations given by
$
	\langle \eta(x,H) \eta(x',H')\rangle
		= \delta^d(x-x') \, G(H-H'),
$
where $G(H-H')$ decays rapidly, approximated
by a delta function for random-field disorder.
The critical exponents at the depinning transition 
have been calculated by $\epsilon$-expansions
\cite{narayan-fisher:1993,nattermann-etal:1992}
and simulations \cite{leschhorn:1993,laszlo,qEW-exponents}.
The upper critical dimension is $d_c=4$, above which
mean-field theory applies \cite{qEW-mft}.
Below we will also discuss so-called columnar noise with
$G(H)\equiv 1$ \cite{parisietpietronero}.

The interface equation~(\ref{eq:qEW}) obeys an invariant 
so that the static response scales as \cite{narayan-fisher:1993}
$\chi(q,\omega=0) \sim q^{-2}$, i.e.,
\begin{equation}
	\gamma/\nu = 2   .
		\label{eq:gamma/nu}
\end{equation}
For forces below $F_c$, the (bulk) response of the interface triggered
by a small increase in $F$ scales as
$	\chi_{\rm bulk} \equiv {d \left< H \right>} / {dF} 
					\sim (F_c-F)^{-\gamma}$.
Right at the critical point one can argue as follows
\cite{narayan-fisher:1993,nattermann-etal:1992}:
the roughness of the interface scales as $\ell^\chi$ and assuming that
$\Delta \left< H \right>$ will scale in the same way it follows 
\begin{equation}
	\gamma = 1 + \chi \nu .
		\label{eq:gamma}
\end{equation}
This yields $\chi + 1/\nu = 2$, i.e., there are only two independent exponents
for depinning described by (\ref{eq:qEW}).
The standard scaling relations are valid for interfaces
with parallel dynamics: all sites with $\partial H / \partial t > 0$
are updated in parallel. 
Note that interfaces with extremal (i.e., one unstable site
at a time) and parallel
drive have the {\it same\/} pinning paths.
This manifests the Abelian character of the LIM
in that the order in which active sites are advanced does not
matter \cite{paczuski-etal:1996}.

\section{Mapping of sandpile dynamics}

Next we will show that the SOC critical behavior can be related exactly to the
slowly driven depinning transition in an interface model.
Thus, Eqs.~(\ref{eq:<s>}) and (\ref{eq:gamma/nu}) are equivalent
and Eq.~(\ref{eq:gamma}) yields an
expression for the correlation length exponent $\nu$ for sandpiles.
The first step is to formulate the stopping of an avalanche in a SOC
system as being due to a pinning path for an interface $H(x,t)$.  
This field is given in the continuum limit by
\begin{equation}
	H(x,t) = \int_{0}^{t} dt' \, \rho(x,t')  ,
			\label{eq:H=int-rho}
\end{equation}
where the order parameter $\rho(x,t)$ is the activity (topplings)
at site $x$ at time $t$, i.e., $\rho=\dot{H}=v\sim f^\theta$.
In words: $H(x,t)$ counts the number of topplings at site $x$ up to time $t$. 
At the microscopic level
this is an exact correspondence between a toppling
and the interface advance. 
A toppling takes place when $z_x > z_c$, which by the relation
\begin{equation}
	z_x = z_c + \frac{\partial H}{\partial t} ,
				\label{eq:z-def}
\end{equation}
yields the dynamics $\partial H/\partial t > 0 ~ \Rightarrow ~ H \to H + 1$,
whereas $H$ is unchanged at the sites where no toppling takes place.
The dynamics of sandpile models thus map to discrete interface equations
where an avalanche takes the interface $H(x,t)$
from one pinning path to the next in the quenched random medium
\cite{paczuski-boettcher:1996,narayan-fisher:1993,%%%%%
zapperi-etal:1998,narayan:1994,cule:1998}. 
Since the interface counts topplings it does not move
backwards and thus Eq.~(\ref{eq:z-def}) effectively reads
${\partial H}/{\partial t}=\theta(z_x-z_c)$, which
is the standard discretization for depinning models \cite{leschhorn:1993}.
We are currently investigating the applicability of such discretization
procedures to various models \cite{allau:1999,unpublished}.

Next, we express $z_x$ in terms of $H(x,t)$ for the specific
models introduced above. The number of grains $z_x$ on site $x$
is $z_{x} = N_{in} - N_{out} + F(x,t)$,
where $N_{in}$ is the number of grains added to this site from 
its $2d$ nearest neighbors (nn) and
$N_{out}$ is the number of grains removed from this site 
due to topplings. The (external) driving force $F(x,t)$ counts
the number of grains added from the outside.
Since $N_{in}=\Sigma_{nn} H(x_{nn},t)$ and $N_{out}=2dH(x,t)$
(for details and extensions to other models see \cite{unpublished})
we arrive at 
\begin{equation}
        \frac{\partial H}{\partial t} = \nabla^2 H - z_c(x,H) + F(x,t) ,
				\label{eq:H-eq}
\end{equation}
where $\nabla^2 H$ is the discrete Laplacian.
The Dirichlet boundary conditions for $z_x$ become $H\equiv 0$
and the dynamics is parallel. Similar connections have been previously
discussed for a charge-density wave model \cite{narayan:1994}
and for a boundary driven rice-pile model \cite{paczuski-boettcher:1996}
(see below).
In the stochastic model, $z_c(x,H)$ is a random variable which
changes after each toppling. 
Thus $z_c(x,H)$ acts like quenched random point-disorder similar to 
$\eta(x,H)$ in Eq.~(\ref{eq:qEW}).
The BTW model has $z_c$ equal to a constant.
The dissipation needed to reach the SOC state (loss of grains $z_x$)
takes place through the BC of $H\equiv 0$.
Using strong boundary pinning may thus give rise to 
the possibility of observing SOC experimentally in 
systems displaying a depinning transition.
We emphasize that the mapping prescription can in principle
be applied to any sandpile model.
For other, more complicated, toppling rules 
\cite{kadanoff-etal:1989,AL:1997} additional terms like
the ``Kardar-Parisi-Zhang'' nonlinearity $|\nabla H|^2$
may appear.

On the internal (fast) time scale the driving force
$F(x,t)$ does not act as a time-dependent noise but
as columnar-type disorder.
It counts all the grains added to the system by the slow drive,
i.e.\ $F(x,t) \to F(x,t) +1$,
and thus increases as function of time in an uncorrelated fashion.
In the opposite limit when a grain is added (e.g.) each time step
(``fast drive'') $F(x,t)$ would correspond to a time-dependent 
noise \cite{narayan-fisher:1993}.
Since $H\equiv 0$ at the boundary and $F$ increases as
function of time the steady-state profile of $H$ will
be close to a paraboloid or, in one dimension, a parabola
(see also \cite{dickman-etal:1998}).
In the steady-state, just after an avalanche, the slowly increasing
force $F$ is balanced by the negative curvature $\nabla^2 H$ of the
paraboloid such that all sites are pinned ($\partial H/\partial t \le 0$).
This illustrates that the interface effectively is driven by a force
equal to the critical force $F_c\equiv \zeta_c-\overline{z_c}$,
where $\overline{z_c}$ is the average of $z_c(x,H)$ in the steady state
(for the BTW model trivially $z_c =2d-1$).
Accordingly, the slow drive reaches the depinning
critical point by adjusting the dissipation to the driving force
such that the velocity (order parameter) is infinitesimal.

The steady-state of the different sandpile models
is described by an equation similar to Eq.~(\ref{eq:qEW}).
Thus the exponent relation (\ref{eq:gamma/nu})
holds and it is equivalent to Eq.~(\ref{eq:<s>}) which
describes the scaling of the average avalanche size (``susceptibility'').
Assuming that a roughness exponent $\chi$ can be defined for sandpile models,
one can argue that Eq.~(\ref{eq:gamma}) is valid also for sandpiles.
Furthermore, the upper critical dimension is $d_c=4$.
Note that the ensuing noise will contain a columnar component
\cite{narayan:1994,cule:1998,parisietpietronero}
due to the random drive $F(x,t)$.
The one-dimensional BTW model has a critical force 
$F_c=1-1=0$, which corresponds to the critical point of the 
columnar-disorder interface model \cite{parisietpietronero}.
In $d>1$, one has $F_c<0$ which in combination with the fact that
the interface by definition cannot move backwards implies that
the BTW model displays a more complicated behavior
than the columnar models investigated in
\cite{narayan:1994,parisietpietronero}. Note also that avalanches in 
stochastic models will have a random structure due to the explicit
point disorder whereas avalanches in the BTW model show a more 
regular behavior \cite{allau:1999,avalanches}.

For the case of the boundary driven one-dimensional
rice-pile models \cite{christensen-etal:1996,AL:1996}
a similar mapping of the dynamics can be done with
an auxiliary field $H(0,t)$ and a drive implemented as
$H(0,t) \to H(0,t) +1$ \cite{paczuski-boettcher:1996}.
The rice-pile models have Dirichlet BC at $x=0$ and Neumann
BC (reflective) at $x=L$ which yields $\left<s\right>\sim L$.
In our picture the boundary drive is $F(1,t) \to F(1,t) +1$ 
and $F(x>1,t)=0$. Because of the Neumann BC 
[$H(L,t)=H(L+1,t)$]  the steady state develops
a parabolic profile with the left branch pointing up
\cite{paczuski-boettcher:1996}.

\section{Various ensembles}

We next consider the more straightforward
cases in which sandpiles are studied
with periodic boundary conditions (amounting to $H(1) = H(L)$
in one dimension). In such cases the SOC steady state can be 
tuned into by various approaches. It can be reached by using a 
carefully tuned bulk dissipation
$\epsilon \sim L^{-2}$ \cite{vespignani-zapperi:1997}.
In this case, periodic BCs are also the best since the scaling of the system
is not a mixture of boundary and bulk scaling \cite{LFH:1998}.
As above, we arrive at
\begin{equation}
        \frac{\partial H}{\partial t} 
		= \nabla^2 H - z_c(x,H) -\epsilon(x,H) + F(x,t)
			\label{eq:H-eq-eps}
\end{equation}
with $H$ periodic. As in Eq.~(\ref{eq:H-eq}), the force $F(x,t)$
is columnar and increases on the slow time scale.
The dissipation $\epsilon(x,H)$ takes now into account
all the grains removed before the site at $x$ topples.
It increases with a (small) probability only when
a site topples and this means that $\epsilon$ explicitly depends on $H$.
Therefore, a dissipation event effectively corresponds to
a shift in the $z_c$ value. Thus, one obtains that the BTW model
with bulk dissipation contains a very weak point-disorder component
(since the increases in $\overline{F}$ equal in the statistical sense
the increases in $\epsilon$). Though point-disorder is in general
expected to be a relevant perturbation, in the infinite system
size limit the Larkin length 
\cite{narayan-fisher:1993,nattermann-etal:1992}
associated with the cross-over
from columnar behavior diverges and thus the avalanche
behavior is not governed by the weak point disorder.
By this argument the BTW models with or without bulk dissipation
are equivalent 
to the same interface depinning equation (\ref{eq:qEW})
in accordance with simulations of the BTW 
and bulk dissipation models \cite{chessa-etal:1998}.
Note that the boundary critical behavior of the BTW model
depends on the specific boundary condition:
Dirichlet BCs display a different behavior
\cite{D=2}, whereas Neumann BCs (reflective)
are similar to the bulk.
In the case of periodic BCs and bulk dissipation,
the $H$-field fluctuates around an average flat profile.
The terms $F(x,t)$ and $\epsilon(x,H)$
will balance each other in the steady state with
an average difference such that $F_c=\zeta_c-\overline{z_c}<0$.
For larger dissipation rates the system moves away from the critical
point and, in analogy to Eq.~(\ref{eq:gamma/nu}),
the bulk susceptibility scales as
$\chi_{\rm bulk} \sim 1/\epsilon \sim \xi_{\epsilon}^{1/\nu_\epsilon}$,
with $\nu_\epsilon=1/2$ \cite{vespignani-zapperi:1997}.

The fixed density (or energy) drive
previously used in simulations \cite{tang:1988,dickman-etal:1998}
corresponds to a normal driven interface. Thus, the situation is such
that $H(x,t=0)=0$, $\zeta = L^{-d} \sum_{x} F(x,0)$ with $F(x,t)=F(x,0)$,
and periodic BCs and $\epsilon(x,H)\equiv 0$ such that no 'grains' are lost.
The control parameter $\Delta=\zeta-\zeta_c$ ($=F-F_c\equiv f$) is varied 
and criticality is only obtained when $\Delta=0$;
note that choosing $\zeta$ corresponds to using a spatially dependent 
force $F(x,0)$ with $\zeta=\left< F(x,0) \right>$. 
Here, the system is not generally
in the SOC steady-state but by letting the control
parameter $\Delta\to 0$ one reaches the critical point
\cite{tang:1988,dickman-etal:1998}.
The noise is set at the beginning
of an avalanche at the columnar values $F(x,0)$.
Depending on the exact nature of the initial configuration 
one may observe a different dynamic behavior but the steady-state 
behavior should
correspond to the slowly driven case \cite{dickman-etal:1998}.

In ``microcanonical'' simulations \cite{chessa-etal:1998prl} one has
dissipation operating on the slow time
scale with exactly the same rate as $F(x,t)$.
Thus microcanonical simulations correspond to
fixed density simulations with a specific initial configuration: 
after each avalanche, the time is reset
to zero, the force is replaced with $F \to F +\nabla^2 H$, and
the forces at $x'$ ($x''$) are increased (decreased)
by one unit where $x'$ and $x''$ are randomly chosen sites.
Finally the interface is initialized, $H\equiv 0$.
Since the $\nabla^2 H$ term does not introduce correlations
this new starting condition is equivalent to the fixed density case
but with the initial configuration chosen to be in the SOC steady state.

Combining the scaling relations (\ref{eq:gamma/nu}) and
(\ref{eq:gamma}) it follows that 
\begin{equation} 
	2+d = D + 1/\nu ,
				\label{eq:scal-rel}
\end{equation} 
where $D = d + \chi$.  In addition,
the average area scales as $\left< \ell^d \right> \sim L^{1/\nu}$.
These relations are also valid
for sandpiles and Eq.~(\ref{eq:scal-rel}) provides estimates for $\nu$: 
in $d=1$, $\nu\approx 1.30$,
and in $d=2$, $\nu\approx 0.78$.
Numerical results yield $\nu=1.25(5)$ ($d=1$, stochastic model) \cite{veje}
and $\nu=0.79(4)$ ($d=2$, BTW model) \cite{dickman-etal:1998}.
Note, however, that the estimates quoted for $\nu$ for sandpile models
depend on the relation $D = d+\chi$, which means that the underlying
assumption is that the roughness exponent $\chi$ can be defined
for slowly driven sandpile models.

\section{Conclusions}

In summary, we have started from the depinning equation
(\ref{eq:qEW}) to discuss the continuum description of self-organized
critical sandpile models. Thus, their upper critical dimension 
is $d_c=4$ and a scaling relation for the correlation
length exponent $\nu$ is obtained.
We find that the BTW model has columnar disorder $F$ on the
avalanche time scale whereas the stochastic models have
 explicitly point disorder included.
Other models with slightly modified toppling rules 
(e.g., the Manna model \cite{manna:1992}) may or may not belong
to the same classes depending on the noise terms arising from
the mapping (this we are currently investigating further
in \cite{allau:1999,unpublished}).
The present approach shows that the relevant noise for sandpiles is 
'quenched'. The physics of sandpiles is such that the random decisions
or events (grain deposition, choices for thresholds) are frozen into
the dynamics of a site as long as it is stable, and their memory
decays only slowly as the activity goes on. 
A recent field theory for 
$\rho(x,t)$ used analogies from systems with absorbing
states and assumed that the noise was
Reggeon field-theory like (i.e., time-dependent
and not quenched) \cite{dickman-etal:1998}.
Physically, the effect which is not incorporated in
such Gaussian correlations is that the pinning forces along
the interface selects a pinning path in the random media
which stops the avalanche.

The mapping between interface and sandpile dynamics allows one
to characterize the sandpile universality classes by
the quenched noise in the interface equations. It also allows
to gain novel insight about the previously introduced ways of reaching
the depinning critical point:
balancing the force with dissipation (slow drive, or self-organized
criticality), tuning the average force (as for fixed density sandpiles),
tuning the interface velocity (extremal drive criticality),
and finally tuning the driving force.
This becomes possible because of the
diffusive character of interface or sandpile dynamics and because of the
Abelian character of the linear interface equation.

%acknowledgements

K. B. L. is supported by the Carlsberg Foundation.

\end{multicols}


\begin{thebibliography}{99}
\bibitem{D=2}
	For the BTW model in $d=2$ avalanches initiated at the 
	boundary have $D=2$ since sites can only topple once.
\end{thebibliography}
\end{document}